\documentclass[rmp,aps,twocolumn,amsmath,amssymb,superscriptaddress]{revtex4-2} 
\usepackage{graphicx}
\usepackage{hyperref}
\usepackage[utf8]{inputenc}
\usepackage[T1]{fontenc}
\begin{document}

\title{Platforms for the realization and characterization of Tomonaga Luttinger liquids}
\author{Isabelle Bouchoule}
\affiliation{Laboratoire Charles Fabry, Institut d’Optique Graduate School, CNRS, Université Paris-Saclay, 91127 Palaiseau, France}
  \author{Roberta Citro}
\affiliation{Physics Department ‘E. R. Caianiello’ and CNR-SPIN, Università degli Studi di Salerno, Fisciano, (SA), Italy.} 
\affiliation{INFN — Gruppo Collegato di Salerno, Salerno, Italy. }
\author{Timothy Duty}
\affiliation{School of Physics, UNSW Sydney, Sydney, NSW 2052, Australia}
\affiliation{Qilimanjaro Quantum Tech, S.L.Carrer de Veneçuela, 7408019 Barcelona, Spain}
\author{Thierry Giamarchi}
\affiliation{DQMP, University of Geneva, 24 Quai Ernest-Ansermet, CH-1211 Geneva, Switzerland}
\author{Randall G.  Hulet}
\affiliation{Department of Physics and Astronomy, Rice University, Houston, Texas 77005, USA} 
\author{Martin~Klanj\v{s}ek}
\affiliation{Jo\v{z}ef Stefan Institute, Ljubljana, Slovenia}
\author{Edmond Orignac}
\affiliation{CNRS, Ens de Lyon,  LPENSL, UMR5672,  F-69342 Lyon, France}
\author{Bent Weber}
\affiliation{School of Physical and Mathematical Sciences, Nanyang Technological University, 21 Nanyang Link, Singapore 637371}

\begin{abstract}
The concept of a Tomonaga-Luttinger liquid (TLL) has been established as a fundamental theory for the understanding of one-dimensional quantum systems. Originally formulated as a replacement for Landau's Fermi-liquid theory, which accurately predicts the behaviour of most 3D metals but fails dramatically in 1D, the TLL description applies to a even broader class of 1D systems,including bosons and anyons. After a certain number of theoretical breakthroughs, its descriptive power has now been confirmed experimentally in different experimental platforms. They extend from organic conductors, carbon nanotubes, quantum wires, topological edge states of quantum spin Hall insulators to cold atoms, Josephson junctions, Bose liquids confined within 1D nanocapillaries and spin chains. In the ground state of such systems, quantum fluctuations become correlated on all length scales, but, counter-intuitively, no long-range order exists. In this respect, this review will illustrate the validity of conformal field theory for describing real-world systems, establishing the boundaries for its application and, on the other side will discuss the spectacular demonstration of how the quantum-critical TLL state governs the properties of many-body systems in one dimension.
\end{abstract}

\flushbottom
\maketitle

\newpage
\tableofcontents

\section{Introduction to Tomonaga-Luttinger liquids}\label{sec:introduction}
While in three dimensional fermionic fluids the low energy excitations can be described in terms of weakly interacting quasiparticles possessing the same quantum numbers as the individual fermions\cite{landau_fermiliquid_theory_dynamics}, such paradigm has been shown to break down in one dimension by S.-I. Tomonaga\cite{tomonaga_model} and J.-M. Luttinger\cite{luttinger_model}. Instead of quasiparticles, the low energy excitations are linearly dispersing collective excitations associated with fluctuations of particle or spin density.\cite{Giamarchi}  The velocity of the particle and spin density modes are usually different leading to the spin-charge separation phenomenon. A striking consequence of the presence of these density modes is that in the ground state, instead of showing a step at the Fermi energy, the energy distribution of fermions shows a power-law singularity.  A similar power law singularity is found in the momentum distribution and in fact all correlation and response functions at momenta integer multiples of the Fermi momentum: one dimensional fermions are in a critical state called \textbf{Tomonaga-Luttinger liquid} (TLL), with critical exponents depending on a single parameter, the so-called Tomonaga-Luttinger parameter.\cite{haldane_xxzchain} At the same time, the compressibility, spin susceptibility and stiffness\cite{kohn_stiffness} are finite and non-singular. The velocities of the collective modes, and the Tomonaga-Luttinger exponents can be expressed as a function of susceptibilities and stiffnesses, allowing a non-perturbative definition. Thanks to the Jordan-Wigner transformation\cite{jordan_transformation}, the Tomonaga-Luttinger  liquid concept is also applicable to one-dimensional quantum antiferromagnets\cite{luther_spin1/2,haldane_xxzchain} and boson systems\cite{haldane_bosons}.  Besides  the original formulation of the Tomonaga-Luttinger liquid concept in the 1980s\cite{luther_spin1/2,haldane_bosons}, chiral Tomonaga-Luttinger liquids\cite{wen_edge_review}, in which particle density modes propagate in a single direction,  and helical Tomonaga-Luttinger liquids\cite{wu_helical_2006,xu_stability_2006} in which counterpropagating spin density modes also carry charge have been introduced to describe edge states respectively in the Fractional Quantum Hall Effect\cite{Klitzing} and in two-dimensional topological insulators. Corrections to the linear dispersion of the modes and their effect on correlation functions is the object of nonlinear Tomonaga-Luttinger liquid theory\cite{imambekov_one-dimensional_2012}.
Initial experimental support for Tomonaga-Luttinger liquid theory was provided by experiments with organic conductors\cite{jerome_quasi_2024}, quantum wires\cite{tarucha_wire_1d,auslaender2005spin}, carbon nanotubes\cite{bockrath_luttinger_nanotubes} and spin chain materials\cite{tennant_kcuf_1d}. New platforms to probe Tomonaga-Luttinger liquid physics have appeared recently with ultracold atomic gases\cite{cazalilla_review_bosons} and Josephson junction chains\cite{fazio_1996,glazman_1997}.

\section{The Tomonaga-Luttinger liquid: formalism}\label{sec:concept}

\subsection{Hamiltonian and observables of the Tomonaga-Luttinger liquid}\label{sec:observables}

For spinless fermions, the Tomonaga-Luttinger liquid Hamiltonian is \cite{haldane_bosons,Giamarchi}
\begin{equation}\label{eq:1component}
    H=\int \frac{dx}{\pi}\left[uK (\partial_x \theta)^2 + \frac u K (\partial_x \phi)^2\right], 
\end{equation}
where $\phi$ and $\theta$ are bosonic fields representing collective excitations of the system — specifically, long-wavelength density and phase fluctuations. They obey the commutation relation $[\phi(x),\theta(x')]=i\pi\Theta_H(x'-x)$,  $\Theta_H$ being the Heaviside step function, while $u$ is the velocity of low-energy excitations and $K$ is the dimensionless Tomonaga-Luttinger liquid parameter.  
The  fermion density is expressed in terms of the operator $\phi(x)$ as  
\begin{equation}\label{eq:spinless-fermion-density}
  \rho(x)=\rho_0 -\frac 1 {\pi}\partial_x\phi + \sum_{ m\in \mathbb{Z}-\{0\}} A_m e^{i 2m (\phi(x)-\pi\rho_0 x)},   
\end{equation}
where $\rho_0$ is the average fermion density, $A_m^*=A_{-m}$, while the fermion annihilation operator is 
\begin{eqnarray}\label{eq:spinless-fermion-annihilation}
    \psi_F(x)=e^{i\theta(x)} \sum_{m\in \mathbb{Z}} B_m e^{i (2m+1)[\phi(x)-\pi\rho_0 x]}. 
\end{eqnarray}
The Jordan-Wigner transformation\cite{jordan_transformation} turns spinless fermion annihilation operators into hard core boson annihilation operators. It yields\cite{haldane_bosons} the expression  
\begin{eqnarray}\label{eq:boson-annihilation}
    \psi_B(x)=e^{i\theta(x)} \sum_{m\in \mathbb{Z}} C_m e^{i 2m [\phi(x)-\pi\rho_0 x)]},  
\end{eqnarray}
for the boson annihilation operator, while~(\ref{eq:spinless-fermion-density}) gives  the boson density. More generally, bosons with repulsive interactions are in a Tomonaga-Luttinger liquid phase\cite{haldane_bosons}. Independently of particle statistics, the charge stiffness\cite{kohn_stiffness} is $\mathcal{D}=uK$ while the compressibility is $\chi=K/(\pi \rho_0^2 u)$, and at low temperature, the specific heat is $C_v=\frac{\pi k_B^2 T}{3u}$. These relations allow a non-perturbative definition of $u$ and $K$.  
Moreover, since hard core bosons can be mapped to spin-1/2 operators\cite{matsubara_hardcore1956}, with $S_j^+\sim (-)^j\psi_B^\dagger(ja)$ and $S_j^z=\rho(x)-1/(2a)$, XXZ spin-1/2 chains also host a Tomonaga-Luttinger liquid phase.\cite{luther_spin1/2,haldane_xxzchain} A fully polarized spin state maps to empty or full Jordan-Wigner fermion band, and a state with partial magnetization corresponds to the partial filling of the fermion band and is a TLL. The role of the compressibility is played by the magnetic susceptibility, $\chi=(g\mu_B)^2K/(\pi u)$ where $\mu_B$ is the Bohr magneton and $g$ the $g$-factor, while charge stiffness is replaced by spin stiffness\cite{shastry_twisted_1990}. The magnetic heat capacity $C_m=\pi k_B^2 T/(3u)$ has a linear temperature dependence, and the Wilson ratio, $R_W=\frac{4}{3}\left(\frac{\pi k_B}{g\mu_B}\right)^2\frac{\chi T}{C_m}=4K$ measures the Tomonaga-Luttinger exponent\cite{Ninios_2012}. 
In the case of spin-1/2 fermions, the two-component Tomonaga-Luttinger liquid Hamiltonian reads 
\begin{eqnarray}
H&=&\sum_{\nu=c,s} H_\nu, \\ 
H_\nu&=&\int \frac{dx}{2\pi} \left[u_\nu K_\nu (\partial_x \theta_\nu)^2 + \frac{u_\nu}{K_\nu} (\partial_x \phi_\nu)^2\right], 
\end{eqnarray}
where the fields satisfy the commutation relation $[\theta_\nu(x),\phi_{\nu'}(x')]=i\pi \delta_{\nu\nu'} \Theta_H(x'-x)$, and $K_s=1$ when the interactions preserve $\mathrm{SU(2)}$ spin rotation symmetry. $H_c$ describes the charge excitations of velocity $u_c$, while $H_s$  describes the spin excitations of velocity $u_s$, showing the spin-charge separation when $u_c \ne u_s$.  
The relations to define non-perturbatively the Tomonaga-Luttinger parameters\cite{schulz_su2} $u_c,u_s,K_c$ involve the charge stiffness $\mathcal{D}_c=2 u_c K_c $, the compressibility $\chi_c=K_c/(\pi \rho_0^2 u_c)$, the spin susceptibility $\chi_s=(g\mu_B)^2/(2\pi u_s)$, and the low temperature specific heat behaves as $C_v(T)=\frac{\pi k_B^2 T}{3u_c}+\frac{\pi k_B^2 T}{3 u_s}$, giving a Wilson ratio\cite{Ninios_2012} $R_W=\frac{2u_c}{u_s+u_c}$.    
The charge density is
\begin{eqnarray}\label{eq:charge-density}
 \rho_c(x)=\rho_0 -\frac {\sqrt{2}} {\pi} \partial_x \phi_c + C \cos (\pi \rho_0 x -\sqrt{2} \phi_c) \cos \sqrt{2} \phi_s+\ldots,    
\end{eqnarray}
the spin density has components 
\begin{eqnarray}\label{eq:spin-density}
\sigma^x(x)&=&\frac{\cos \sqrt{2} \theta_s \cos \sqrt{2} \phi_s}{\pi\alpha} + C' \cos (\pi \rho_0 x -\sqrt{2} \phi_c) \cos \sqrt{2} \theta_s, \\
 \sigma^y(x)&=&\frac{\sin \sqrt{2} \theta_s \cos \sqrt{2} \phi_s}{\pi\alpha} + C' \cos (\pi \rho_0 x -\sqrt{2} \phi_c) \sin \sqrt{2} \theta_s, \\
 \sigma^z(x)&=& -\frac 1 {\pi\sqrt{2} } \partial_x \phi_s + C' \cos (\pi\rho_0 x -\sqrt{2} \phi_c) \sin \sqrt{2} \phi_s,
\end{eqnarray}
with $\alpha$ a short distance cutoff, and the annihilation operators for particles of spin $\sigma$ and momentum near $r \pi\rho_0/2$ ($r=\pm 1$) 
\begin{eqnarray}
\psi_{r,\sigma} =\frac 1 {\sqrt{2\pi\alpha}}  \exp\left\{i\left[\frac{\theta_c - r\phi_c}{\sqrt{2}} +\mathrm{sign}(\sigma)  \frac{\theta_s - r\phi_s}{\sqrt{2}} \right] \right\},   
\end{eqnarray}
with $\psi_{\sigma}(x)=\sum_r e^{i r \pi \rho_0 x/2} \psi_{r,\sigma}(x)$. At half filling, $\rho_0=1/a$ replacing the operator $\phi_c$ with $0$ in the spin density~[] Eq.(\ref{eq:spin-density})] gives the bosonized expression of the spin operators in the spin-1/2 Heisenberg chain. 
In the helical liquid\cite{wu_helical_2006,xu_stability_2006}, spin and Fermi momentum are locked together, making the Tomonaga-Luttinger liquid single component [Eq.~(\ref{eq:1component})] as in the case of spinless fermions. The fermion annihilation operators
reduce to 
\begin{eqnarray}\label{eq:ti-fermions}
\psi_{+,\uparrow}=\frac{e^{i (\theta -\phi)}}{\sqrt{2\pi\alpha}}, \; 
\psi_{-,\downarrow}=\frac{e^{i (\theta +\phi)}}{\sqrt{2\pi\alpha}}, 
\end{eqnarray}
and the expression of the charge density is unchanged with respect to spinless fermions. However, the spin density and the spin raising operators are:
\begin{equation}
\psi_{+,\uparrow}^\dagger\psi_{+,\uparrow} -\psi_{-,\downarrow}^\dagger  \psi_{-,\downarrow} = \partial_x \theta/\pi, \hspace{0.2truecm} \psi^\dagger_{+\uparrow} \psi_{-\downarrow} = \frac{e^{i 2\phi}}{2\pi\alpha}. 
\end{equation}

\subsection{Ground state correlation functions}

The Tomonaga-Luttinger liquid is characterized by a power-law decay of
the ground state correlation function relating immediately the
Luttinger liquid parameters to experiments, as we will discuss below.
In a single component Tomonaga-Luttinger liquid, the ground state
correlation functions $R_{n,m}^{n',m'}(x,t)$ behave as\cite{Giamarchi}
\begin{widetext}
\begin{eqnarray}\label{eq:gs-corr}
\langle e^{i [n \theta(x,t)+m\phi(x,t)]}  e^{-i [n'\theta(0,0) + m' \phi(0,0)]} \rangle = \delta_{n,n'}\delta_{m,m'} \left(\frac{\alpha}{\alpha+i (ut-x)}\right)^{\frac 1 4\left(\frac{n}{\sqrt{K}} - m\sqrt{K} \right)^2} \left(\frac{\alpha}{\alpha+i (ut+x)}\right)^{\frac 1 4\left(\frac{n}{\sqrt{K}} + m\sqrt{K} \right)^2}, 
\end{eqnarray}
\end{widetext}
where $\alpha$ is a short distance cutoff. The power-law singularities in Eq.~(\ref{eq:gs-corr}) are determined by $n,m$ and the Tomonaga-Luttinger liquid exponent $K$.  From the correlation functions above, one finds the generalized susceptibilities $\chi_{n,m}(q,\omega)$ at frequency $\omega$ and wavevector $q$ using Fourier transformation. They present power law singularities\cite{Giamarchi}  
\begin{eqnarray}
\chi_{n,m}(q,\omega) &\sim&
                            \left[\frac{(\omega-uq)\alpha}{u}\right]^{\frac
                            1 4\left(\frac{n}{\sqrt{K}} + m\sqrt{K}
                            \right)^2 -1 } \nonumber \\
                     &\times& \left[\frac{(\omega + u q)\alpha}{u}\right]^{\frac 1 4\left(\frac{n}{\sqrt{K}} - m\sqrt{K} \right)^2 -1 },  
\end{eqnarray}
so that for $\omega=r u q (r=\pm)$, a divergence is obtained when the exponent $(n K^{-1/2}+r m K^{1/2})^2/4 -1$ is negative, and a cusp singularity (along with a finite contribution) when it is positive. The former case\cite{Giamarchi} is realized with charge density wave ($n=0,m=2$) response functions in the repulsive case ($K<1$) or superconducting ($n=2,m=0$) response functions in the attractive case. The latter case is realized in the case of the momentum distribution of spinless fermions ($\omega=0,n=1,m=\pm 1$), resulting in the absence of a step at the Fermi energy.     
In two-component Luttinger liquids, the correlation function factorizes into the product of a spin and a charge correlation function, both of the form~(\ref{eq:gs-corr}), with $u,K$ replaced by respectively $u_\sigma,K_\sigma$ and $u_\rho,K_\rho$. In the ground state, the Fourier transform is expressible\cite{iucci_fourier_2007} in terms of the Appell hypergeometric functions\cite{olver_nist_2010}, and power law singularities now appear when $\omega=\pm u_{c,s} q$. The ground state spectral functions of spin-1/2 fermions\cite{meden_spectral1992,voit_spectral1993,orignac_spectral2011} are also expressed using the Appell functions. Green's functions of fermions in position-frequency space\cite{Braunecker2012} are expressed in terms of modified Bessel functions\cite{olver_nist_2010}.

\subsection{Thermal correlation functions}

At positive temperature, the correlation functions in the single component Tomonaga-Luttinger liquid become\cite{Giamarchi}
\begin{widetext}
\begin{eqnarray}\label{eq:thermal-corr}
&& R_{n,m}^{n',m'}(x,t,T) = \langle e^{i [n \theta(x,t)+m\phi(x,t)]}  e^{-i [n'\theta(0,0) + m' \phi(0,0)]} \rangle \nonumber \\ &&=  \delta_{n,n'}\delta_{m,m'} \left(\frac{\frac{\pi k_B T \alpha}{u}}{\sinh\left(\frac{\pi k_B T}{u} [\alpha+i (ut-x)]\right)}\right)^{\frac 1 4\left(\frac{n}{\sqrt{K}} - m\sqrt{K} \right)^2} \left(\frac{\frac{\pi k_B T \alpha}{u}}{\sinh\left(\frac{\pi k_B T}{u} [\alpha+i (ut+x)]\right)}\right)^{\frac 1 4\left(\frac{n}{\sqrt{K}} + m\sqrt{K} \right)^2}, 
\end{eqnarray}
\end{widetext}
and  their Fourier transforms are now expressed using the Euler Gamma function\cite{schulz_quantum_1983,Giamarchi}. This result demonstrates that the TLL state is quantum critical as its energy scale is defined solely by temperature.
Since the correlation functions decay exponentially over the thermal length $u/(\pi k_B T)$, for $\omega, uq \ll k_B T$ their Fourier transforms become  Lorentzians while for $\omega, uq \gg k_B T$, the ground state power law behavior is preserved.
Then, from a simple scaling analysis 
   $ \chi_{nm}(q,\omega, T)\sim max[u \delta q, \omega, k_B T]^{\nu-2}$
where the exponent is $\nu=(n^2/K+m^2 K)/2$.
In Tomonaga-Luttinger liquids of spin-1/2 fermions, no closed form analytic expression is known, but spectral functions at finite temperature have been calculated numerically\cite{nakamura_spectral1997}.
The tunneling density of states and the tunneling differential conductivity  can still be obtained analytically\cite{bockrath_luttinger_nanotubes}
\begin{equation}\label{eq:didv-tll}
 \frac{dI}{dV} = I'_0 T^{\alpha} \left|\Gamma\left(\frac{1+\alpha} 2  +i \frac{eV}{2\pi k_B T} \right)\right|^2 \cosh \left(\frac{eV}{2k_B T}\right), 
\end{equation}
with $I_0$ a non-universal prefactor, and $\alpha=(K_\rho+K_\rho^{-1}-2)/4$ in the case of fermions with spin and $\alpha=(K+K^{-1}-2)/2$ in the helical liquid. In a chiral Tomonaga-Luttinger liquid, the theory predicts\cite{wen_edge_review,Chang2003} for a fractional filling $\nu=1/(2m+1)$ an exponent $\alpha=2m$.


\section{Itinerant condensed matter TLLs} \label{sec:Josephson}

Itinerant condensed matter systems, such as Tomonaga-Luttinger liquids (TLLs), represent a fascinating class of one-dimensional quantum systems characterized by their distinct properties from traditional Fermi liquids. Unlike conventional three-dimensional metals, where electron interactions can often be treated as perturbative, TLLs exhibit strong correlations that fundamentally alter their behavior.  The theoretical framework of TLL provides a robust description of these phenomena. It applies not only to quantum wires\cite{tarucha_wire_1d,auslaender2005spin,jompol2009,steinberg2008,deshpande_electron_2010} and carbon nanotubes\cite{bockrath_luttinger_nanotubes,zhao_nanotube_2018} but also to edge states\cite{wen_edge_review} in the fractional quantum Hall effect and certain quasi-one-dimensional\cite{watson_multiband_2017,jerome_quasi_2024} conductors, highlighting the universal nature of these one-dimensional systems in condensed matter physics. One specific example of itinerant condensed matter TLL is the 1D array of Josephson junctions.

\subsection{Josephson junction chains }

Theoretically, it was realized some time ago that the low-energy Hamiltonian of a 1D array, or chain, of Josephson junctions coincides with that of a bosonic Tomonaga-Luttinger liquid with backscattering interactions\cite{fazio_1996,glazman_1997}. In a clean system, one expects such interactions to lead to a superfluid-insulator transition, where the insulating states are distinguished by Mott lobes associated with the filling fraction of extra Cooper pairs. However, the simple picture is complicated by the existence of fabrication inhomogeneities and the ubiquitous so-called ``offset charge'' disorder of the small superconducting islands comprising the chain. The nature of the insulating state can be viewed as a competition between commensurate and incommensurate pinning. The offset charge disorder in small superconducting islands is ``maximal'', meaning that it varies roughly uniformly in the interval [-1,1] in units of the primary charge. The result is a Luttinger mode with completely incommensurate disorder in the backscattering, which exhibits an insulating 1D Bose glass\cite{giamarchi_1988,fisher_bosons_localization} for sufficiently strong interactions.

In the absence of quantum fluctuations, the system would be analogous to the classical pinning of an elastic chain.\cite{fukuyama_1978} Such a chain is characterized by a localization length known as the Larkin length, $N_\mathrm{L}$. Classically, one would need to apply a force, here an external voltage, to initiate a flow of current. For chains larger than $N_\mathrm{L}$, the pinning force is the critical voltage divided by the number of junctions in the chain, and has dependence, $e V_\mathrm{c}/N \propto  N_\mathrm{L}^{-2} \propto E_\mathrm{0}^{ -1/3}W^{4/3}$, where $E_\mathrm{0}$ is the elastic energy, and W is the width of the disorder, equal to the Bloch bandwidth of the junction energy band for maximal disorder. We note $W \sim \hbar\omega _{p}$, a characteristic energy of the junctions. In order to experimentally compare chain families of widely varying $\hbar  \omega _{p}$, and chain length $N$, we express the critical voltage and Bloch bandwidth as dimensionless variables, $v \equiv e V_\mathrm{c}/N \hbar  \omega _{p}$, and $w \equiv W/\hbar  \omega _{p}$. In the classical limit, $v =a w^{4/3}$, with prefactor $a =b \left (K/\Lambda \right )^{1/3}$, $b$ a constant $\mathcal{O} \left (1\right )$, and $\Lambda$ the electrostatic screening length.

\begin{figure*}
	\includegraphics[width=1\linewidth]{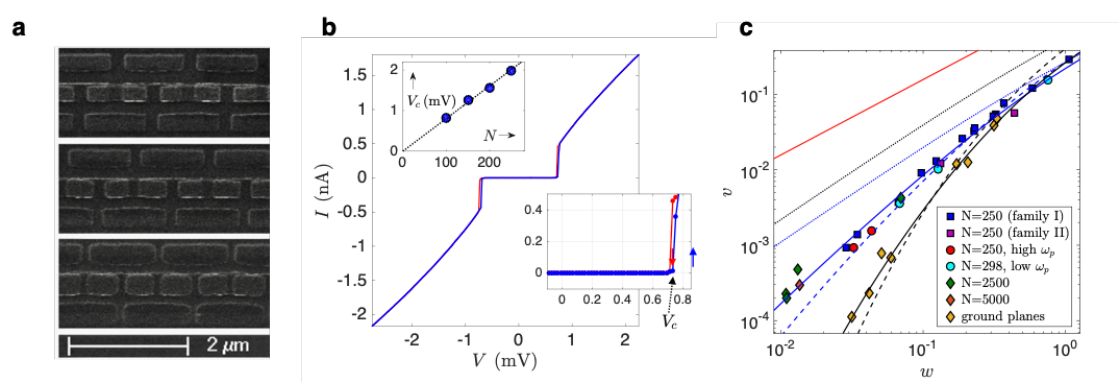}
	\caption{{\label{figJJ} \bf Josephson junction chains as pinned TLL’s.} {\bf a,} SEM~micrograph showing a family of chains with nominal junction size 300$ \times $400 nm. The specific capacitance is 54 fF/$\mu$m\textsuperscript{2}. The precise junction area is modulated by the EBL exposure dose. {\bf b,} (main plot) Experimental determination of the critical voltage for a chain with 250 junctions. The blue data is obtained upon stepping up from zero voltage, and the red when stepping back down from non-zero current. (lower inset) Close up of the small voltage region where the critical voltage is extracted.  A very small hysteresis region is present for this device. The critical voltage is taken to be the voltage having maximum $dI/dV$ upon stepping up from zero voltage. (Upper inset) Linear dependence of $V_c$ on chain length, $N$. {\bf c,} Scaled critical voltage $v$, versus scaled Bloch bandwidth $w$. Symbols represent different `families' distinguished by plasma frequency $\omega_p$, length and presence of ground plane. The solid red line is theory for independent QPS across each junction and no disorder and has slope = 1. Solid lines are the quantum theory of a disordered T-L mode with fitted values of screening length $\Lambda$\,=\,13.1 (blue), and $\Lambda$\,=\,4.0 (black), respectively. These exhibit a $w$-dependent slope, 4/(3\,-\,2$K$), where $K(w)$ is the Tomonaga-Luttinger parameter. For small $w$, $K\propto \Lambda ^{ -1}  \ln w$:  for the same range of $w$, $K$ is enhanced by the decreased screening length of devices with ground planes, resulting in a stronger departure from the classical result. The dotted lines show the classical depinning result, slope = 4/3, for $\Lambda$\,=\,13.1 (dotted blue) and $\Lambda$ = 4.0 (dotted black). Also plotted for comparison are the quantum results for  $\Lambda$\,=\,7.7 (blue dashed), and $\Lambda$\,=\,3.2 (black dashed) using screening lengths inferred from the gate-dependent periodicity of $dI/dV$ at large $w$ and biases $V>V_c$. The figure is reproduced from Ref.~\cite{Cedergren_etal-2017}.
	}
\end{figure*}

The effect of quantum fluctuations, as described by the Tomonaga-Luttinger theory, is to increase the localization length\cite{suzumura-loc,giamarchi_1988} with increasing T-L constant,
$K$, such that $N_\mathrm{L} \propto w^{ -2/\left (3 -2 K\right )}$, which diverges at the Bose glass-superfluid (BG-SF) transition, $K_\mathrm{c} =3/2$. The critical voltage then scales as $v =a w^{\alpha }$, where $\alpha  =4/(3 -2 K)$. The dominant effect for $K \neq 0$ is apparent in the exponent $\alpha $. The prefactor $a$ is only weakly dependent on $K$. The dependence of the critical voltage on scaled bandwidth $w$, T-L constant $K$, and other parameters have been experimentally determined by fabricating and measuring a large ensemble of Al/AlO$_ {x}$/Al single-junction chains as shown in Fig.~\ref{figJJ}. By systematically varying $K$ through $w$ and $\Lambda$, we confirm Tomonaga-Luttinger behavior in Josephson junction chains.

\section{Tomonaga-Luttinger liquid physics in spin chains and ladders} \label{sec:spin-chains}

Magnetic insulators containing localized exchange coupled spins comprising chains or ladders offer a particularly convenient platform for the realization of TLL physics. They feature well defined and relatively simple Hamiltonians amenable to theoretical treatment and are experimentally accessible both by powerful probes of spin dynamics, like neutron scattering and nuclear magnetic resonance (NMR), and by bulk magnetic and thermodynamic properties measurements.

\begin{figure*}
	\includegraphics[width=1\linewidth]{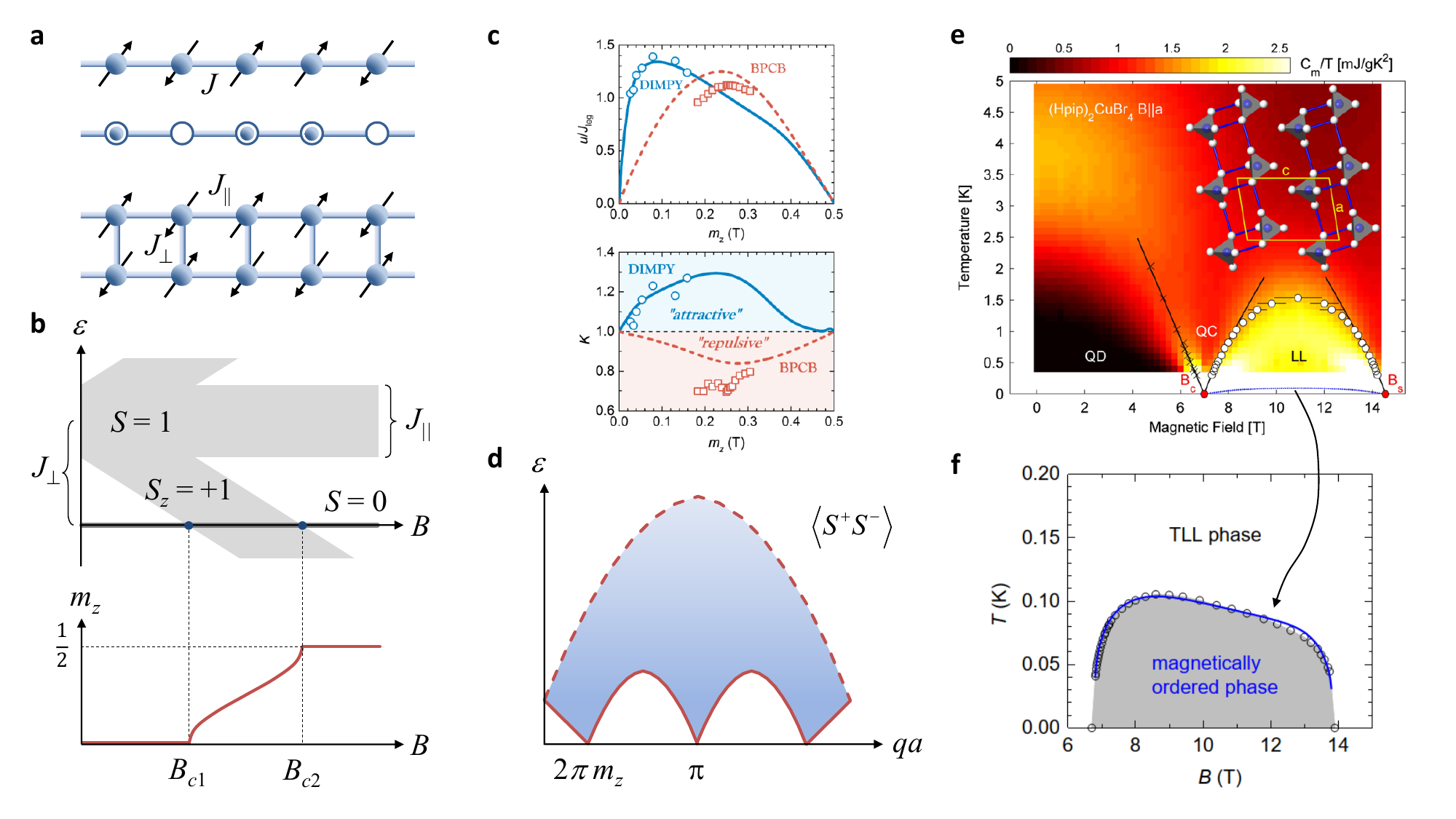}
	\caption{{\bf TLL in spin chains and ladders.} {\bf a,} Spin chain with exchange coupling $J$, its mapping to the system of spinless fermions (spin up/down maps to the presence/absence of fermion), and spin ladder with exchange couplings $J_\perp$ along the legs and $J_\parallel$ along the rungs. {\bf b,} Singlet $S=0$ and triplet $S=1$ energy levels in a spin ladder, and its normalized magnetization $m_z$ per spin as a function of the magnetic field $B$. {\bf c,} TLL parameters $u$ (divided by $J_\parallel$) and $K$ as a function of $m_z$ in strong-rung or repulsive ($J_\perp/J_\parallel/=3.6$, dashed red line) and strong-leg or attractive ($J_\parallel/J_\perp=1.7$, solid blue line) regimes, together with the data for BPCB (squares) and DIMPY (circles), respectively (the figure is taken from Ref.~\cite{Jeong_2016}). {\bf d,} Schematic intensity of the transverse magnetic excitations in a spin chain as a function of energy $\varepsilon$ and wave number $qa$. {\bf e,} Phase diagram of the spin ladder material (C$_5$H$_{12}$N)$_2$CuBr$_4$ (BPCB) as a function of the magnetic field and temperature exhibits a non-magnetic quantum disordered (QD) phase of spin singlets, a TLL phase and the region of quantum criticality (QC) in between (the figure is taken from Ref.~\cite{Ruegg_2008}). The inset shows the crystal structure of the material. {\bf f,} The boundary of the low-temperature magnetically ordered phase indicated in {\bf e} measured by NMR (circles) and described in the model of weakly coupled TLLs (blue line). The figure is reproduced from Ref.~\cite{Klanjsek_2008}.
	}\label{figCL}
\end{figure*}

Spin chains and ladders exhibit the following paradigmatic physics. Whereas a Heisenberg antiferromagnetic spin chain is characterized by a single exchange coupling $J$ between neighboring spins (Fig.~\ref{figCL}a), there are two couplings characterizing a Heisenberg spin ladder, $J_\perp$ along the legs and antiferromagnetic $J_\parallel$ along the rungs of the ladder (Fig.~\ref{figCL}a). In zero magnetic field, the $S=1/2$ spin ladder is in a non-magnetic $S=0$ singlet ground state. The lowest lying spin excitations form a $S=1$ triplet band of width $\sim J_\parallel$ separated by a gap of $\sim J_\perp$ from the ground state. As shown in Fig.~\ref{figCL}b, increasing magnetic field $B$ along $z$ linearly decreases the gap of the lowest lying triplet band with $S_z=+1$. At a lower critical field $B_{c1}$, the gap closes, the spin ladder enters a gapless state and starts to get magnetized. This state persists up to the upper critical field $B_{c2}$, at which the ladder gets fully magnetized as the whole triplet band sinks below the energy of the singlet state, so that the gap between both reopens. A similar behavior is found in spin chains with bond alternation and in gapped Haldane $S=1$ spin chains.\cite{zapf_BEC_review} The low-energy behavior in all these cases is essentially equivalent to the one in $S=1/2$ antiferromagnetic spin chain where the roles of the $S_z=\pm 1/2$ states is played by the singlet and the lowest lying triplet~\cite{Giamarchi_1999}. Furthermore, as the $S_z=\pm 1/2$ states can be mapped to the occupied and empty states in the canonical spinless fermion chain using Jordan-Wigner transformation~\cite{Giamarchi}, spin chains and ladders apparently also exhibit TLL physics in the gapless phase between the critical fields. When discussing their physics, we frequently use the fermion language originating from this transformation.

At the edges of the gapless phase corresponding to the bottom and the top of the Jordan-Wigner fermion band, the velocity $u$ of spin excitations starts from zero and reaches the maximum at an intermediate field, where the dispersion of the fermion band is the steepest (Fig.~\ref{figCL}c). Meanwhile, the interaction parameter $K$ takes the non-interacting value $1$ at the critical fields where the fermion band is nearly empty or nearly full with low density of particles or holes. Away from the critical fields, the value below or above $1$ respectively signifies repulsive or attractive interactions between spin excitations~\cite{Giamarchi_1999} (Fig.~\ref{figCL}c). Gapless spin excitations appear only at a few points in the reciprocal space~\cite{chitra_spinchains_field}, roughly corresponding to the wave vectors across the Fermi surface in the fermionic picture: at $qa=\pi$ and $2\pi m_z$ or $qa=0$ and $\pi(1-2m_z)$ for transverse or longitudinal spin excitations (i.e., perpendicular or parallel to the magnetic field), respectively, where $a$ is the lattice constant~\cite{Giamarchi}. At higher energies, the $S=1$ triplet excitations, fractionalized into pairs of $S=1/2$ spinons, form the continuum schematically shown in Fig.~\ref{figCL}d for the transverse case, whose analytical expression at low energies is known~\cite{Giamarchi}. The analytical expressions for the spin-spin correlation functions are given by Eq.~(\ref{eq:thermal-corr}) with $n=1$, $m=0$ or $n=0$, $m=2$ multiplied by some non-universal and unknown amplitudes~\cite{Giamarchi}. The dependence of these and of the TLL parameters $u$ and $K$ on the normalized magnetization $m_z$ per spin and thus on the magnetic field are then calculated by fitting the analytical expressions to the numerical results\cite{bouillot_dynamical_ladder} obtained for the corresponding spin model using density matrix renormalization group (DMRG) calculations~\cite{Hikihara_2001}.

In the following, we describe the experiments establishing TLL physics in material realizations of spin chains and ladders. The first material allowing this is KCuF$_3$ containing parallel antiferromagnetic $S=1/2$ chains of Cu$^{2+}$ spins~\cite{Lake_2005}. A large exchange coupling $J/k_B=390$~K limits the study to the vicinity of $m_z=0$ corresponding to a half filled fermion band. A measurement of the intensity of transverse magnetic excitations using inelastic neutron scattering confirms the predicted continuum in Fig.~\ref{figCL}d (with $m_z=0$). Precise data further allow to show that the intensity around $qa=\pi$ in the energy range up to $\sim J$ scales with the ratio of energy and temperature, and that this scaling fits perfectly to the analytical TLL prediction~\cite{Lake_2005}. The first thermodynamic test of the field induced TLL behavior was made using Ni(C$_9$H$_{24}$N$_4$)NO$_2$ClO$_4$ (abbreviated as NTENP)~\cite{Hagiwara_2006}. NTENP contains parallel bond-alternating antiferromagnetic chains of Ni$^{2+}$ $S=1$ spins where the exchange coupling along the chains alternates between the values $J/k_B=54.2$~K and $\delta J/k_B=24.4$~K, yielding a singlet dimer ground state with a gap towards the triplet excitations. This gap closes at the critical field $B_{c1}=12.4$~K parallel to the chains, where the material enters a low-temperature magnetically ordered phase extending up to the temperature of the order of $1$~K. The magnetic specific heat at several magnetic fields above this phase exhibits the predicted linear temperature dependence. Its slope, which is predicted to scale with $u^{-1}$, is indeed found to decrease above $B_{c1}$ in accordance with increasing velocity $u$ and is almost perfectly reproduced by numerical diagonalization of the spin Hamiltonian. However, this study touches only the bottom of the fermion band.

To probe the whole fermion band, the material with attainable upper critical magnetic field is needed. The first such material was (C$_5$H$_{12}$N)$_2$CuBr$_4$ (abbreviated as BPCB) with $B_{c1}\approx 7$~T and $B_{c2}\approx 14$~T. The material contains parallel ladders of Cu$^{2+}$ $S=1/2$ spins with $J_\perp/k_B=12.9$~K and $J_\parallel/k_B=3.6$~K, hence in the strong-rung regime~\cite{Klanjsek_2008}. The phase diagram of BPCB as a function of the magnetic field and temperature (Fig.~\ref{figCL}e) is mapped using the magnetic heat capacity~\cite{Ruegg_2008}. Just like in NTENP, the TLL phase is characterized by the linear temperature dependence of the magnetic specific heat. As regards spin dynamics, the low-energy transverse magnetic fluctuations at $qa=\pi$ (Fig.~\ref{figCL}d) as monitored by NMR perfectly match the TLL prediction $1/T_1 \sim T^{1/(2K)-1}$ in the whole range between $B_{c1}$ and $B_{c2}$, corresponding to the whole fermion band~\cite{Klanjsek_2008}. As the ladders in BPCB are weakly coupled, these fluctuations lead to the appearance of magnetically ordered phase at temperatures below $\sim 0.1$~K (Fig.~\ref{figCL}f). An unusual asymmetric shape of the phase boundary as a function of the magnetic field obtained by NMR or neutron diffraction is perfectly reproduced in a model of coupled TLLs where the coupling is treated in the mean-field approximation~\cite{Klanjsek_2008,Thielemann_2009_1}. And finally, inelastic neutron scattering allows to directly observe the fractionalization of $S=1$ triplons into $S=1/2$ spinons across $B_{c1}$ and to confirm the predicted intensity of transverse magnetic excitations (Fig.~\ref{figCL}d) with incommensurate gapless points at $qa=2\pi m_z$ and $2\pi (1-m_z)$~\cite{Thielemann_2009_2}.

A related material (C$_7$H$_{10}$N)$_2$CuBr$_4$ (abbreviated as DIMPY) allows to probe the opposite, strong-leg regime of spin ladders~\cite{Hong_2010} realizing a rare example of attractive interaction between spinon excitations~\cite{Schmidiger_2012}. The material contains parallel ladders of Cu$^{2+}$ $S=1/2$ spins with $J_\perp/k_B=9.5$~K and $J_\parallel/k_B=16.5$~K~\cite{Schmidiger_2012}. Above a lower critical field of $B_{c1}=3.0$~T, weakly coupled spin ladders magnetically order below $\sim 0.3$~K~\cite{Ninios_2012,Schmidiger_2012,Jeong_2013}. Like in BPCB, the boundary of the ordered phase mapped by magnetic specific heat can be perfectly reproduced in a model of weakly coupled TLLs~\cite{Schmidiger_2012}. Above this boundary, the TLL behavior is again confirmed by the observed linear temperature dependence of the magnetic specific heat $C_m$~\cite{Hong_2010}. In addition, the Wilson ratio $R_W$ is found to be close to the theoretically predicted value of $4K$ where the interaction TLL parameter is in the attractive regime $K>1$~\cite{Hong_2010}. The parameter $K$ can also be determined directly from the power-law behavior of the transverse spin-spin correlation function at $qa=\pi$ using NMR, where the power is a simple function of only $K$, and is indeed found to be above $1$~\cite{Jeong_2013}. A combination of both TLL parameters can as well be determined less reliably from both the magnetic specific heat and magnetic susceptibility using the corresponding TLL expressions containing $u$ and $K$ given in \ref{sec:observables}~\cite{Jeong_2016}. As shown in Fig.~\ref{figCL}c, the obtained data for BPCB and DIMPY agree relatively well with the theoretical prediction and nicely demonstrate the difference between repulsive ($K<1$) and attractive ($K>1$) regimes~\cite{Jeong_2016}. Finally, the limits of the low-energy TLL description are demonstrated in a high-precision inelastic neutron scattering study of DIMPY~\cite{Schmidiger_2013}.

A TLL regime of strongly repulsive interactions with $K<1/2$ is found in $S=1/2$ spin chains with XXZ exchange anisotropy where the ratio of exchanges along $z$ and in the $xy$ plane is $\Delta>1$~\cite{Okunishi_2007}. Parallel chains of this kind are realized by Co$^{2+}$ spins in the material BaCo$_2$V$_2$O$_8$ where $\Delta\approx 2$~\cite{Kimura_2007}. This anisotropy leads to antiferromagnetic ordering in each chain, which is suppressed at the critical magnetic field $B_c=3.9$~T along $z$, above which the chain enters a TLL state~\cite{Kimura_2008_1}. The interaction TLL parameter $K$ is predicted to vary from the lowest possible value $1/4$ at $B_c$ to the non-interacting value $1$ at the saturation magnetic field $B_s=22.8$~T applied along $z$~\cite{Okunishi_2007,Takayoshi2018}. This is experimentally confirmed through the observed power-law temperature dependence of the longitudinal spin-spin correlation function in NMR~\cite{Klanjsek_2015}. Namely, in the range $K<1/2$, longitudinal magnetic excitations in a spin chain at an incommensurate wave number $qa=\pi (1-2m_z)$ become dominant over the transverse ones at $qa=\pi$ detected in the previously described cases~\cite{Okunishi_2007}. This leads to the unusual incommensurate magnetic ordering of weakly coupled spin chains at low temperatures~\cite{Kimura_2008_1}, whose nature is confirmed by inelastic neutron scattering~\cite{Kimura_2008_2}. Similar results are obtained in the related compound SrCo$_2$V$_2$O$_8$, where the antiferromagnetic ordering is suppressed at $B_c=4$~T, and a TLL behavior is demonstrated by NMR above the ordering temperature~\cite{cui_field-induced_2022}.

\section{Edge states and hinge states of topological matter} \label{sec:topological-insulators}

\subsection{1D electronic structure in 2D topological insulators}

\begin{figure}[ht]
\centering
\includegraphics[width=0.9\linewidth]{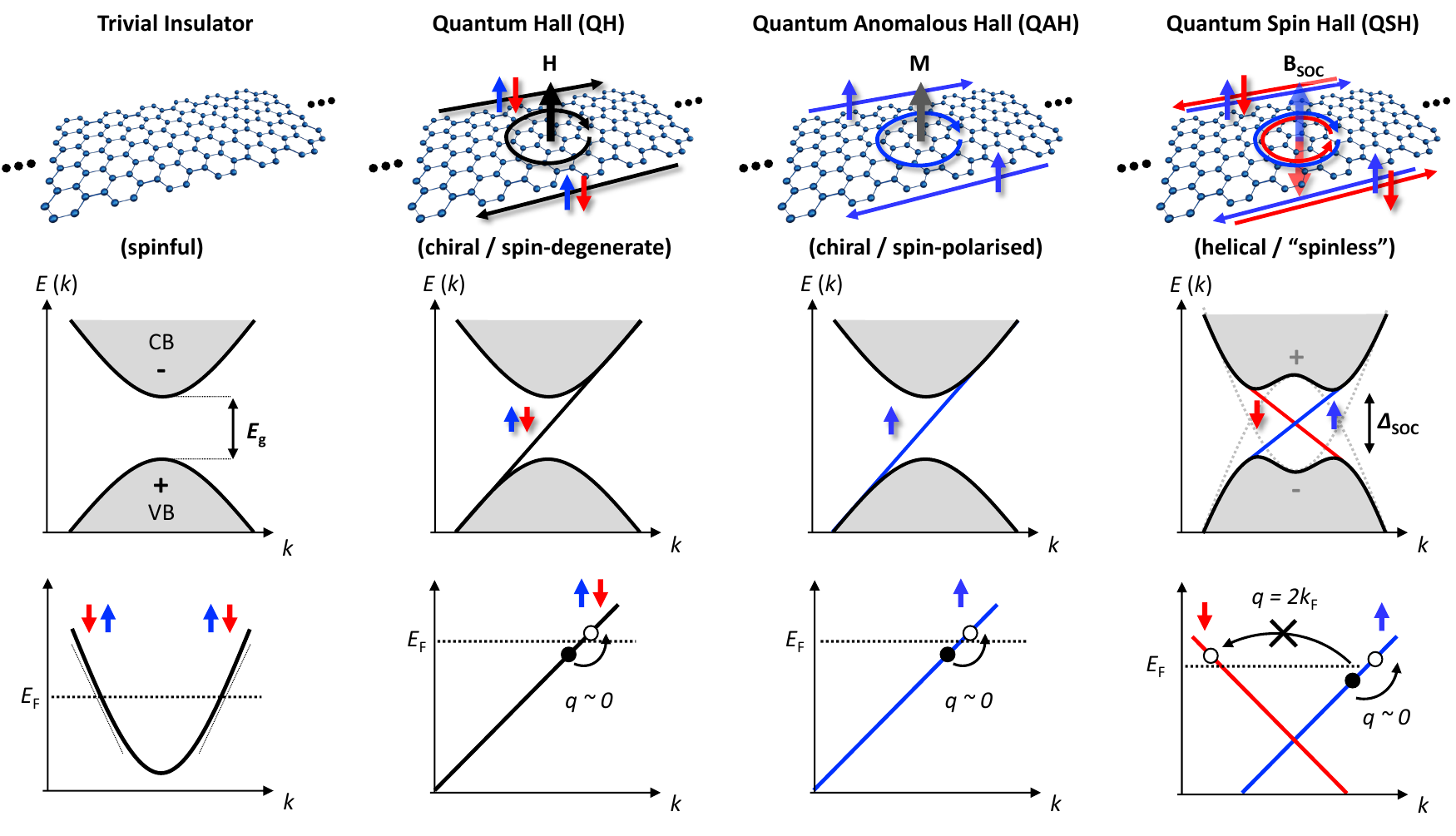}
\caption{\textbf{TLLs in the edge states of 2D topological insulators (TIs).} Realizations of 1D electronic structure and Luttinger liquids at the boundaries of different members of the 2D TI family. Different from trivial insulators in which conduction (CB) and valence (VB) bands are separated by the band gap, 1D metallic edge states persist at the crystalline boundaries of different TIs whose properties (\emph{chiral} vs. \emph{helical}, \emph{spinful} vs. \emph{spinless}) depend on the property of the 2D bulk they surround. The different electronic dispersions governing bulk and edge and scattering processes therein (and their suppression) are indicated.}
\label{fig:TI-1}
\end{figure}

A natural realization of 1D electronic structure arises within the boundary modes of 2D topological insulators (TIs) (Fig.~\ref{fig:TI-1}). Different from conventional electrical insulators (or semiconductor) in which conduction and valence bands are strictly separated by the electronic band gap, topological insulators possess a band-gap only in the bulk of the crystal while metallic boundary states persist on their surfaces and edges. 

Regardless of dimensionality, what makes topological boundary modes fundamentally different from their trivial counterparts (such as e.g. Shockley \cite{Shockley} or Tamm\cite{tamm_1932} states) is that (i) their existence is ensured by topological invariants of the bulk band structure -- a principle called \emph{bulk-boundary correspondence} \cite{bulk-boundary} -- and that (ii) these systems often carry (at least approximately) linearly dispersing 1D Dirac fermions at their boundaries that are either chiral or helical, spin-degenerate or spinless, depending on which class of topological insulator is being considered (Fig.~\ref{fig:TI-1}d).

Different kinds of 2D topological insulators \cite{Klitzing} are classified by their topological invariants -- integer numbers characterizing the electronic band structure that is robust upon adiabatic (smooth) deformations. One of these invariants, the Chern number $n \in \mathbb{Z}$, quantifies the total Berry flux through the Brillouin zone and thus the total number of 1D boundary modes. Crossing a topological phase boundary with some system parameter requires a gap closing and reopening, giving rise to a specified number of 1D modes at the edge.

The first-known 2D topological phase of matter is the quantum Hall effect (QHE) \cite{Klitzing1980}, later put into the context of band topology by Thouless \cite{Thouless1982}. In the QHE, a strong magnetic field $H$ externally applied perpendicular to the plane of a 2D electron gas forces electrons into closed orbits leading to the opening of a gap in the 2D bulk density of states (Landau quantization). Gapless \emph{chiral} edge states\cite{Wen1990} persist at the boundary, spin-degenerate at low magnetic field (Fig.~\ref{fig:TI-1}b). 

Haldane \cite{Haldane1988} later discovered that a similar situation can be realized in a hexagonal lattice of antiferromagnetic order, in which the crystal's intrinsic magnetization $M$ can take the role of an external magnetic field. This gives rise to the quantum anomalous Hall (QAH) state \cite{Haldane1988} that corresponds to topological edge states that are also \emph{chiral}, i.e. \emph{spin-polarized} with only one spin orientation present. 

Different from both the QHE and QAH states, in the quantum spin Hall (QSH) insulator \cite{KaneMele, HBZ},  the spin-orbit interaction acts as an \emph{effective} magnetic field whose polarity depends on the spin of the electron. The QSH state can be realized from an inversion of electronic bands of different parity (Fig.~\ref{fig:TI-1}d) whereby the spin-orbit coupling (SOC) opens a gap at the band crossing points. As such an ``inverted'' band gap needs to close at the border to a surrounding trivial vacuum with non-inverted bands (Fig.~\ref{fig:TI-1}a), gapless pairs of spin-polarized boundary modes arise that are \emph{helical} with spin locked to the crystal momentum such that spins of different polarity travel in opposite direction along the QSH boundary. The QSH state has been reported in semiconductor heterostructures with inverted bandstructures such as HgTe/CdTe \cite{Konig2007, Strunz2019} and InAs/GaSb \cite{Li2015}, as well as more recently in a range of atomically-thin crystals \cite{Lodge2021, 2024Roadmap}, such as the group-IV \cite{Bampoulis2023} and group-V \cite{Stuehler2020TomonagaLuttinger} Xenes, the transition metal dichalcogenides (TMDCs) \cite{Fei2017, Tang2017, Wu2018, jia_2022}, and 3D Dirac semimetal thin films \cite{collins2018electric}.

As shown above, the different members of the 2D TI family provide a vast playground for the realization of 1D electronic structure with different manifestations of Luttinger Liquids -- both \emph{chiral} \cite{Wen1990, wen_edge_review} and \emph{helical} \cite{wu_helical_2006,citro2013,hsu_helical_2021} -- \emph{spinful} or \emph{spinless}. In the 1D chiral and helical edges scattering is strongly suppressed (often referred to as ``topological protection'' of electronic states). In particular, back-scattering \cite{Giamarchi} is absent in chiral edge states, due to a lack of final states, while the helical edge states of QSH insulators are protected from back-scattering by spin conservation. 

\subsection{Experimental demonstrations of chiral and helical TLLs}

A significant amount of experimental work has focused on investigating Luttinger liquids within the chiral edges of quantum Hall states (see Refs.\cite{Chang2003, Fujisawa2022} for comprehensive reviews). Indeed, quantum Hall states offer a considerable degree of tunability \cite{Fujisawa2022}, depending on the electronic filling factor $\nu$ determining the number of chiral edge states. Possible scenarios include co-propagating pairs of edge modes with opposite spin ($\nu=2$), spin-polarized single modes ($\nu=1$), fractional modes, and even counter-propagating (spin-degenerate or non-degenerate) modes of adjacent quantum Hall domains separated by a thin gate-electrode \cite{Kamata2014, Prokudina2014_Tunable, Fujisawa2022}. TLLs have just recently been detected in the helical edge states of QSH insulators \cite{Li2015, Stuehler2020TomonagaLuttinger, jia_2022, Strunz2019}, while no Luttinger liquid claims have been made, to our knowledge, for the chiral edges of QAH (or Chern) insulators \cite{Yu2010, Chang2013}. %

\begin{figure}[ht]
\centering
\includegraphics[width=1\linewidth]{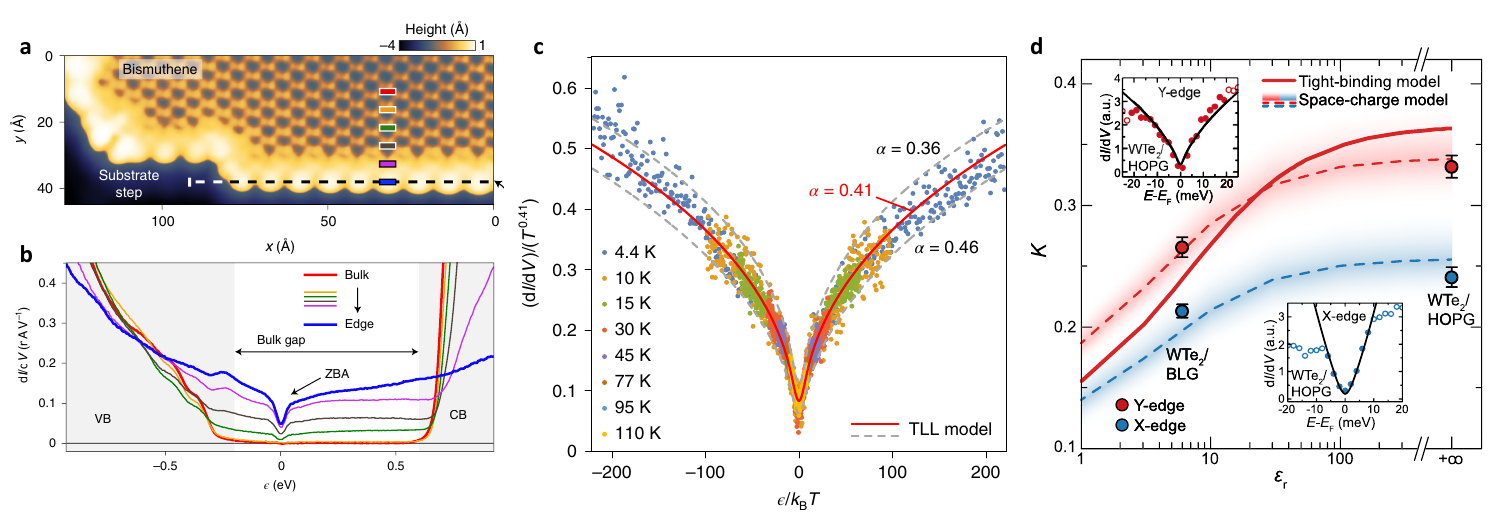}
\caption{{\bf Experimental demonstrations of TLLs in 2D TIs.} Luttinger liquids in the helical edges states 2D topological (quantum spin Hall) insulators as probed by scanning tunneling microscopy and spectroscopy (STM) at cryogenic temperature. {\bf a,} Atomic structure of hexagonal bismuthene monolayer, visualized by STM. {\bf b,} Spatial dependence of tunneling spectra of the 2D bulk at varying distance from the 1D edge as indicated in {\bf a}. A suppression of the tunneling density of states (DOS) into the edge state at at $E_{\rm F}$ (ZBA) is clearly visible. {\bf c,} Universal scaling of the tunneling DOS in temperature and bias voltage used to extract the Luttinger parameter $K$. {\bf d,} Tunability of the Luttinger parameter as a function of the effective dielectric constant of the substrate material in the 2D TI tungsten ditelluride (WTe$_2$). Panels {\bf a-c} were taken from Ref.~\cite{Stuehler2020TomonagaLuttinger}. Panel {\bf d} was taken from Ref.~\cite{jia_2022}.}  
\label{fig:TI-2}
\end{figure}

Similar to early investigations in 1D wires\cite{deshpande_electron_2010} and quantum Hall edge states \cite{Chang1996QH, Chang2003}, experimental probes of chiral and helical Luttinger liquids in 2D TIs have largely focused on tunnelling spectroscopy by either electron transport or local probe spectroscopy. The smoking gun signature for a TLL is thereby the presence of a pseudogap in the measured tunneling density of states at the Fermi level that, according to Eq.~(\ref{eq:didv-tll}), scales as a power-law, universally, in bias voltage and temperature. Curiously, quantum Hall liquids at bulk filling factor $\nu > 1$ have been reported consistent with  Fermi liquid behavior\cite{Hilke2001FQH}. Moreover, for $\nu<1$, the universal tunneling exponent was not related to the bulk filling factor\cite{Chang2003}.  
While several experimental reports on 2D TIs have shown the presence of a pseudogap features within the DOS of the topological edge states at the Fermi level \cite{Tang2017, Fei2017, Reis2017, collins2018electric, Stuehler2020TomonagaLuttinger, jia_2022, Que2023}, consistent with a Luttinger liquid, only few examples exist \cite{Li2015, Stuehler2020TomonagaLuttinger, jia_2022} in which the universal scaling dependence has been systematically investigated. The first report of universal scaling for tunneling into the helical edge states of a QSH insulator was demonstrated for inverted InAs/GaSb heterostructures, in transport spectroscopy \cite{Li2015}. Fig.~\ref{fig:TI-2} shows more recent demonstrations for the atomically-thin QSH insulators \cite{Lodge2021} bismuthene \cite{Stuehler2020TomonagaLuttinger} and WTe$_2$ \cite{jia_2022}, both probed in scanning tunneling microscopy (STM). Both these studies have allowed for precise extractions of the Luttinger parameters $K = 0.42$ \cite{Stuehler2020TomonagaLuttinger} and $K = 0.2 \sim 0.4$ \cite{jia_2022, Que2023}, respectively, the latter tunable via both dielectric screening  \cite{jia_2022} and field-effect doping \cite{Que2023}. \\
An interesting feature, common to the tightly confined ($\xi < 1$~nm) topological edge modes of atomically-thin QSH systems \cite{Stuehler2020TomonagaLuttinger, jia_2022}, are their strong electronic interactions \cite{Bieniek2023Theory}, reflected in Luttinger parameters $K < 1/2$ \cite{Stuehler2020TomonagaLuttinger} or even $K < 1/4$ \cite{jia_2022}. While such strong electronic interactions push the perturbative models to their limits, they promise a unique testbed for the discovery of new physics. The presence of strong electronic interactions in helical edge liquids can have relevant consequences. On the one hand, they can facilitate undesired scattering channels \cite{Maciejko2009_Kondo} that may affect the stability of the QSH state \cite{xu_stability_2006} possibly leading to a breakdown of the much-coveted ``topological protection'' \cite{Moore2023_breakdown}. On the other hand, strongly interacting helical edge liquids \cite{hsu_helical_2021} -- when combined with superconducting pairing -- promise a platform for the realization of non-Abelian quasiparticles beyond Majorana fermions. 

\subsection{Inter-edge Tunneling Device}
Inter-edge tunneling in Luttinger liquids has gathered significant attention for its role in edge state dynamics in quantum Hall systems \cite{Fujisawa2022}. Inter-edge tunneling refers to the process where electrons tunnel between different edge states, which can profoundly affect the transport properties of the system. This phenomenon is particularly interesting in fractional quantum Hall states, where the tunneling can reveal information about the fractional charge and anyonic statistics of the quasiparticles involved. Theoretical studies, such as those by Wen\cite{wen1992} and Kane and Fisher\cite{Kane1992}, have shown that the tunneling current in these systems follows a power-law behavior as a function of temperature and bias voltage, with exponents that depend on the Luttinger liquid parameters. These insights have been pivotal in understanding the non-trivial correlation effects and the emergent exotic physics in low-dimensional quantum systems. More recent experimental approaches have considered devices enabling scattering and inter-edge tunneling \cite{Chamon2009_corner-junction} of edge modes in narrow channels  \cite{Prokudina2014_Tunable} or quantum point contact (QPC) constrictions \cite{Rokhinson_FQH, Cohen2022}. 

\begin{figure}[ht]
\centering
\includegraphics[width=0.7\linewidth]{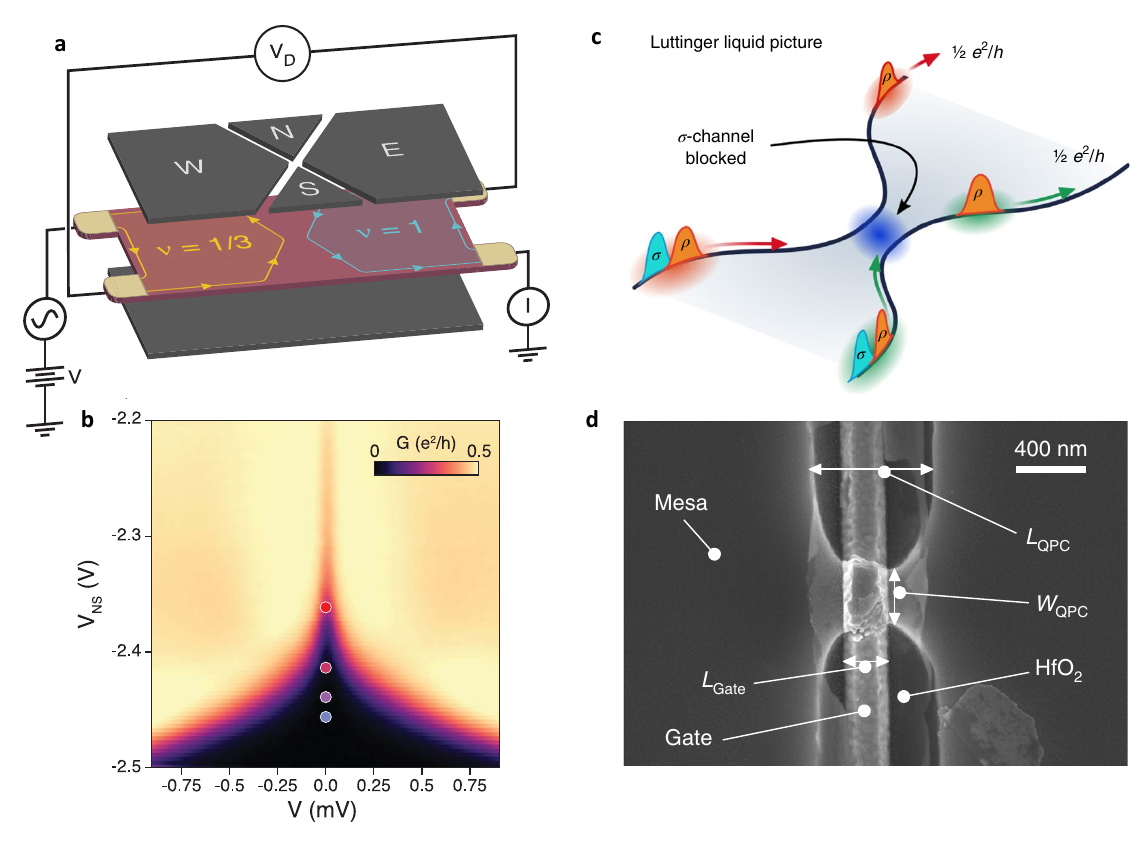}
\caption{{\bf Realizations of inter-edge tunneling devices to probe TLLs at the edges of quantum Hall and quantum spin Hall insulators. a-b,} Tunneling spectroscopy of a graphene point contact connecting integer ($\nu = 1$) and fractional ($\nu = 1/3$) quantum Hall states. Panel {\bf b} shows the finite bias spectroscopy of the QPC conductance and its characteristic zero-boas suppression. {\bf c-d,} Point contact constriction, lithographically defined in the quantum spin Hall regime of an inverted HgTe/CdTe heterostructure. Panels {\bf a-b} were taken from Ref.~\cite{Cohen2022}. Panels {\bf c-d} were taken from Ref.~\cite{Strunz2019}.}
\label{fig:TI-3}
\end{figure}

While the bulk of experimental demonstrations to date have focused on inter-edge tunneling devices in the (integer and fractional) quantum Hall regimes \cite{Fujisawa2022, Rokhinson_FQH, Cohen2022}, recent work also includes point contact spectroscopy measurements of helical QSH edges \cite{Strunz2019}. Other proposals to measure the interaction strength in helical liquids include microwave impedance measurements \cite{Bocquillon2020_microwave}. Examples of devices enabling inter-edge tunneling are provided in Figure\ref{fig:TI-3}. 

Other one-dimensional (1D) topological edge modes representing a fascinating area of condensed matter physics, are the valley-Hall boundary modes in graphene and hinge-modes in higher-order topological insulators (HOTIs). In graphene, valley-Hall boundary modes arise due to the breaking of inversion symmetry, leading to edge states localized at the boundaries of the material that are protected by the valley degree of freedom \cite{MacDonald}. Similarly, HOTIs host 1D hinge-modes, which are robust against perturbations due to their topological nature. These modes emerge at the intersections of the material's surfaces, as seen in both experimental and theoretical studies \cite{Drozdov2014, Jack2019, Wang2023}. These edge and hinge states exemplify how topology can lead to protected modes that have potential applications in next-generation electronic devices.

A further addition to the family of 2D TIs are fractional topological phases \cite{Stern2016} as recently realized in Moir\'e materials \cite{Andrei2021} created from twisted layers of van-der-Waals semiconductors \cite{Cai2023, Xu2023, Kang2024}. Similar to the fractional quantum Hall (FQHE) liquids, these systems carry fractionally charged excitations due to strong electronic correlations. However, different from the FQHE effect, these excitations coexist with non-trivial band topology at zero magnetic field. This makes them highly sought-after phases of matter for the realization of proposals for topological quantum computing. Both fractional QAH \cite{Cai2023, Park2023, Xu2023} and fractional QSH \cite{Kang2024} states have so far been realized.

\section{Realization in cold atom platforms}\label{sec:cold atoms}

\subsection{Pinning transition in 1D Bose gases}

The Tomonaga-Luttinger Liquid framework describes not only 1D fermionic
gases but it also
applies to the low energy description of 1D bosonic gases.
One-dimensional Bose gases are obtained in cold atoms setups, using
either a 2-dimensional deep optical lattice that confines the atoms in each tube
of the lattice or a  micro-magnetic trap to freeze out the transverse degree of
freedom.
In many experiments, the gas is free along the longitudinal direction up to
a slowly varying longitudinal potential, whose effect can be captured within
a local density approximation. Thus, the system is described locally by
a Galilean invariant system, in which case the Luttinger liquid parameter
is linked to the speed of sound and density according to $K= \pi \rho_0/ v$.
Moreover, in most experiments the interactions between Bosons is well
modeled by a contact repulsive two-body interaction of strength $g$, whose
relative strength is quantified by the dimensionless parameter
$\gamma = m g/(\hbar^2 \rho_0)$. The 
system then realizes the famous Lieb-Liniger model
and $K$ can be computed exactly as a function of $\gamma$: it goes to 1 as $\gamma$ goes to infinity,
which corresponds to the hard-core Boson limit (Tonks gas), and it takes the asymptotic value
$K\simeq \pi/\sqrt{\gamma}$ for $\gamma\rightarrow 0$.  The realization of a Tonks-Girardeau gas, as demonstrated in the seminal works by Weiss et al.\cite{kinoshita2004} and Bloch et al.\cite{paredes2004}, has provided a profound experimental platform for exploring one-dimensional quantum systems, while Schmiedmayer et al.\cite{Hofferberth2007} further advanced the field by observing finite-temperature correlation functions of a Luttinger liquid through interference experiments, offering critical insights even in regimes with very high Luttinger parameters, i.e. $\gamma\rightarrow 0$.
Long-wavelength physics, obtained ignoring the oscillating terms in
Eq.\ref{eq:spinless-fermion-density}, reduces to the physics of phonons, or
sound waves, and many experimental
results obtained in 1D Bose gases can be understood
within this framework, both at equilibrium and out-of-equilibrium.

On the other hand, oscillating terms in Eq.~\ref{eq:spinless-fermion-density}
play an important role  in the presence of an external potential whose wavelength is close
to the inter-particle distance $1/\rho_0$. The Hamiltonian then takes the form of
a Sine-Gordon Hamiltonian. This model presents a quantum phase transition from a
delocalized state characterized by algebraically decaying correlation functions to
a localized state with a first order equal-time correlation function decaying exponentially.
In the case of a periodic potential whose wavelength is exactly $1/\rho_0$, this transition
occurs when the periodic potential amplitude exceeds a critical value $V_c$, which depends on
$K$.  Remarkably, $V_c$ vanishes when approaches $K_c$: for $K<K_c$ a periodic potential of
vanishing amplitude is sufficient to localize the system.
This transition is of Kosterlitz-Thouless type, the 2D physics arising here thanks to the
equivalence between a quantum field 1D problem and a 2D thermal classical field,
and the Luttinger parameter at the transition is  $K_c=2$. Those predictions have been
verified in cold atoms experiments~\cite{haller_pinning_2010,Boeris2016}. 

\subsection{Spin-charge separation in 1D atomic fermions}

Unlike three-dimensional (3D) systems whose low-energy excitations are fermionic quasi-particles, the low-energy excitations of one-dimensional (1D) fermions are collective bosonic spin- and charge-density waves (SDW/CDW) that disperse linearly, as described by the Tomonaga-Luttinger liquid (TLL) theory\cite{luttinger_model,tomonaga_model,Giamarchi}. A peculiar result is that the SDW and the CDW of an interacting 1D Fermi gas propagate at different speeds, thus showing a spatial separation of spin and charge excitations in the gas. Recently, a series of experiments with ultracold atoms in an optical lattice was performed on a single-site resolved 1D Hubbard chain, leading to the observation of the fractionalization of spin and charge quantum numbers\cite{hilker2017}, including the modification of the SDW wavevector by density-doping and by spin-polarization\cite{salomon2019}, and the study of simultaneous spin and charge dynamics after performing a quench \cite{vijayan2020}. Quite remarkably these studies allowed to see directly the non-local string structure predicted by the theory between the spinons and holons. However, the excitations produced by the quench are not in the linearly dispersing regime, so they can only be qualitatively be compared with the TLL theory\cite{Bernier_2014}. 
 
\begin{figure}[ht]
\centering
\includegraphics[width=0.5\linewidth]{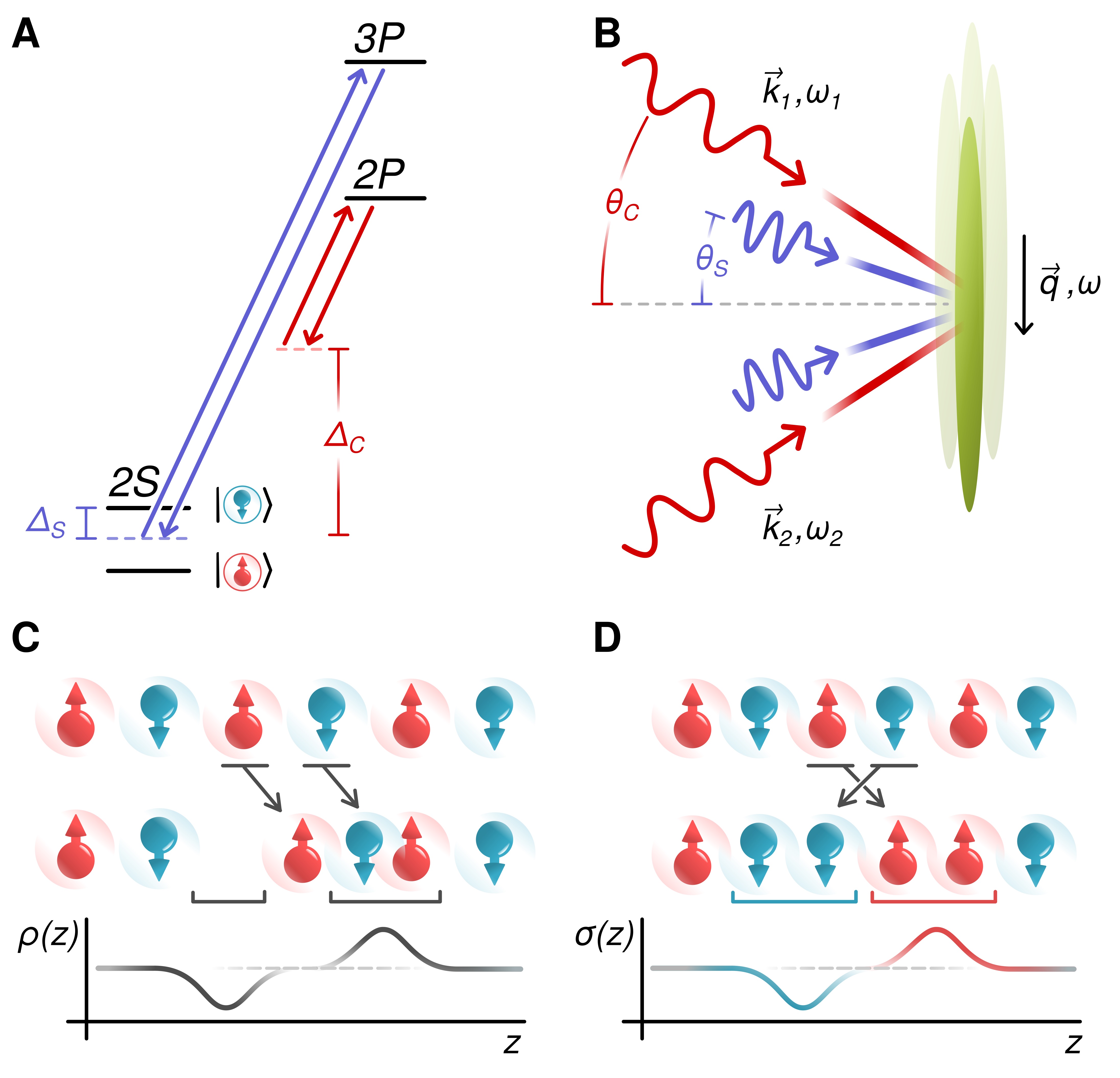}
\caption{Spin and charge excitations via Bragg spectroscopy. (A) Relevant transitions and laser detunings for spin (S, violet)
 and charge (C, red) excitations in $^6$Li atoms. (B) Bragg beams (1, 2) oriented to produce a standing wave along the tube axis. A momentum transfer
 $\vec{q} = \vec{k_1}-\vec{k_2}$ is imparted to the atoms. (C,D) Schematic diagram of the charge and spin excitations, showing an excitation of a particle-hole pair, or a
 spinon pair. At the bottom is shown the effect on the total density $\rho(x)$
 and spin density $\sigma(x)$. The figure is reproduced by Ref.\cite{hulet_2022}
}
\label{fig:Bragg}
\end{figure}

More recently, the collective low-energy excitation spectra inherent to spin-charge separation have been measured using Bragg spectroscopy\cite{hulet_2022} as depicted in Fig.\ref{fig:Bragg}. This method uses two laser beams with frequency difference $\omega$ intersecting at an angle that determines the momentum $q$ imparted by a Bragg scattering event. For a system composed of a balanced mixture of two spin components, there are two contributions, $S_{\sigma,\sigma'}(q,\omega)$, to the dynamical structure factors (DSF's), where
$S_{\sigma \sigma'}(q,\omega)=\frac{1}{2\pi} \int_{-\infty}^{\infty} F_{\sigma \sigma'}(k,t)e^{i\omega t}$, and where $F_{\sigma \sigma'}(k,t)=\frac 1 N <\rho_{k \sigma}(t) \rho_{-k \sigma'}(t)>$. The charge- and spin-density DSF's are then given by:
 \begin{equation}
 S_{C,S}(q,\omega)=2\lbrack S_{\uparrow,\uparrow}(q,\omega) \pm S_{\uparrow,\downarrow}(q,\omega) \rbrack
\end{equation}
 The DSF's, $S_{C,S}(q,\omega)$, are proportional to the total momentum transferred to the atoms by the Bragg pulse, $P(q,\omega)$, a quantity that may be directly measured using time-of-flight imaging. As shown in Fig.\ref{fig:Bragg}A, the excitation of a CDW requires a symmetric detuning $\Delta_C$ of the Bragg beams from two-photon resonance, while the SDW is excited with an antisymmetric detuning $\Delta_S$. In both cases, the magnitude of the detuning from resonance with the excited state must be sufficiently large to minimize the rate of spontaneous emission compared to the coherent Bragg scattering process. In the case of the SDW, this criterion was met by using the ultraviolet 2S-3P transition, rather than the technically simpler red 2S-2P transition. The momentum transfer is:
 $P(q,\omega)=\lbrack  S(q,\omega)-S(-q,-\omega)=S(q,\omega)(1-exp(-\hbar \omega/k_BT)$.
 
 The Bragg spectroscopy is performed by applying the pair of Bragg beams to the $^6$Li atoms. The intensity of the beams are adjusted to limit the loss of atoms from to spontaneous scattering
 and to ensure that the momentum transfer is in the linear response regime for either mode over the entire range of interaction strengths. Following 
 the Bragg pulse, the atoms are released from the lattice, and after time-of-flight 
 are imaged with a phase-contrast method following time-of-flight. The Bragg signal is proportional to the number of out-coupled atoms and the DSF may be extracted for both charge and spin. Bragg
 spectroscopy may additionally be extended beyond the linear regime where studies of the NLL provides an opportunity to benchmark novel calculations, including
band curvature effects and spin-charge\cite{imambekov_one-dimensional_2012,hulet_2022,Cavazos_2023}.

\section{Beyond the Tomonaga-Luttinger liquid physics} \label{sec:other-platforms}

\subsection{The contribution of nonlinearities}

The TL liquid theory has served as a useful paradigm for describing one-dimensional (1D) quantum fluids in the limit of low energies. This theory is based on a linearization of the dispersion relation. In the phenomenological LL approach,
the conventional justification for such simplification is irrelevance of the nonlinearity in the renormalization group sense. Progress in understanding 1D quantum fluids beyond the low-energy limit has been proposed \cite{imambekov_one-dimensional_2012}, and studied experimentally and compared with theory\cite{barak_interacting_2010,moreno_nonlinear_2016,jin_momentum-dependent_2019,vianez2022,hulet_2022,Cavazos_2023}. The irrelevant
 terms hardly affect the fermion propagator away from the singular lines in space-time, $x\pm vt \gg
 \sqrt{t/m*}$. The vicinity of these lines defines the singular behavior of the spectral function. It has been shown for all spinless 1D fermionic models with short range
 interactions that the single-particle spectral function at $p << k_F$ is universal. In the vicinity of the Fermi
 wave vector $+k_F$ and $p,\omega > 0$, the spectral function  $A(p,\omega)$ is a universal function of a single argument:
\begin{equation}
\label{eq:spectral_nl}
    A(p,\omega)= A(\epsilon), \epsilon=\frac{\omega-v p}{p^2/2m*},
\end{equation}
 where $p$ is measured from the closest Fermi point, and $\hbar=1$. The function $A(\epsilon)$ is very different from the LL theory predictions, yet it depends only on the LL parameter K. The asymptotic behaviour of $A(\epsilon)$ for $\epsilon >> 1$ recovers the LL theory predictions, but at $|\epsilon \pm 1| \ll 1$ the spectral function is described by a power-law with new exponents\cite{tsyplyatyev2015,tsyplyatyev2016}. The exponents are different from the predictions of the LL theory yet can still be analytically expressed in terms of $K$. The exponent can be determined numerically by the nonlinear dynamics in nonequilibrium Fermi gases\cite{Abanin2004,bettelheim2006_hydodynamics,Bettelheim2008}.
 
The methodology described above, which combines perturbation theory relative to interaction strength, and innovative techniques for handling finite-size properties of integrable models, is applicable to a broad spectrum of 1D fluids, including 1D spin liquids, electrons in quantum wires\cite{tsyplyatyev2015,tsyplyatyev2016}, and cold atoms confined in 1D traps. Such an approach additionally demonstrates how nonlinear dispersion leads to finite particle lifetimes and influences the transport properties of 1D systems at finite temperatures.

\subsection{Coupled chains}
In bosonic chains in the TL liquid regime the addition of interchain hopping or interchain interaction induces respectively superfluid  or density wave long range order at low temperature\cite{cazalilla_coupled_XY_cold}. For spin chains and ladders coupled by exchange interaction, the analog is the formation of a Néel order with XY or Ising symmetry.\cite{Giamarchi_1999,Orignac_2000,bouillot_dynamical_ladder,zapf_BEC_review} In both cases, the ordering is described by chain mean-field theory\cite{Giamarchi_1999,cazalilla_deconfinement_long,katanin_magnetic_2007}, and the pretransitional fluctuations by the random phase approximation\cite{Giamarchi_1999,bocquet02_chain_rpa,dupont_dynamical_2018,horvatic_direct_2020}. In coupled fermionic chains, the situation is more subtle\cite{bourbonnais_couplage,boies_couplage}. Although superfluidity is favored at strong attraction, and density wave ordering at strong repulsion, at weak or intermediate interaction strength, the interchain hopping $t_\perp$ drives a crossover to the Fermi liquid state below a coherence temperature $T_{coh.}\sim E_F (t_\perp/E_F)^\frac{1}{2-(K_c+K_c^{-1}+K_s+K_s^{-1})}$.\cite{boies_couplage} In the presence of strong forward scattering, it has been proposed that a sliding Tomonaga-Luttinger phase could be stabilized\cite{emery_smectic}, however in the case of purely repulsive interaction with interchain hopping this is limited to a narrow region of parameter space\cite{vishwanath_slide_LL} inaccessible with screened Coulomb interactions\cite{fleurov_instability_2018}. 

\subsection{ Beyond Luttinger-Liquid paradigm: Generalized Hydrodynamics }
The LL model accounts for the physics of the system only as long as one considers
low energy physics. Capturing the dynamics of states very far from the ground state is {\it a priori}
a very hard task. However, remarkably, in the case of integrable 1D systems an exact description exists in
the asymptotic case of slow and long length scale dynamics, which is the Generalized Hydrodynamic (GHD) theory.
This theory has been tested experimentally in
cold-atoms setups implementing the Lieb-Liniger model~\cite{schemmer_generalized_2019,malvania_generalized_2021,doyon_generalized_2023}. The basic idea is that
integrable systems hold an infinite number of conserved quantities.
These local conserved quantities involve the notion of rapidities, which are
velocities that label the eigenstates of the 1D integrable system
and whose number is equal to the number of particles. For slow and long length scale dynamics, one can assume that the gas
is described by its rapidity distribution, which is a 
two-dimensional function $\rho(x,v)$. For each small interval around $v$,
$\int dx \rho(x,v)$ is a conserved
quantity which implies that, for any $v$, $\rho(x,v)$ obeys a continuity equation.
The flux term, to order in $1/L$, where $L$ is the typical length scale of the dynamics, 
is simply a functional of the local rapidity distribution and it is  written as
\begin{equation}
  j(v,x) = v_{\rm{eff}[\rho]}(v) \rho(x,v)
\end{equation}
where, at any position $x$, $v_{\rm{eff}[\rho]}$ is a functional of the local rapidity distribution
whose exact calculation uses the Bethe-Ansatz machinery~\cite{CastroAlvaredo2016,Bertini2016}.
 The effect of an external smoothly varying
potential $V(x)$ introduces a flux along the direction $v$. Finally,  the function $\rho(x,v)$
obeys a Liouville-like equation, which is nothing else than the Euler-scale GHD equation~\cite{CastroAlvaredo2016,Bertini2016}
\begin{equation}
  \frac{\partial}{\partial t} \rho +\frac{\partial }{\partial x} \left ( v_{\rm{eff}} \rho \right ) -\frac{\partial V}{\partial x}
  \frac{\partial }{\partial v} \rho=0.
  \label{eq:GHD}
\end{equation}
The LL theory is recovered as the limit of GHD if one consider only small
perturbations of the ground state. Drude weights\cite{kohn_stiffness} have been measured in one-dimensional quantum gases\cite{schuttelkopf_2024} and were found to agree with GHD predictions.

\subsection{Quantum critical Luttinger liquid behavior and spinon confinement}

 Low dimensional quantum magnets are interesting systems because of the emerging collective
 behavior arising from strong quantum fluctuations which can lead to LL behavior. The one-dimensional (1D) S=1/2 Heisenberg antiferromagnet is a paradigmatic example whose excitations, known as spinons\cite{faddeev_spin1/2}, carry spin $S=1/2$, half the spin of a magnon. In the Luttinger liquid phase, these fractional modes are deconfined and have linear dispersion\cite{bernard_spinons_1994,bouwknegt_spinons}. They can be reconfined by the
 application of a staggered magnetic field or a dimerization\cite{haldane_dimerized}. Spinons and their confinement  have been observed in a variety of experimental systems\cite{kenzelmann_breather,kenzelmann05_spinon_bs,lake_finiteT_spinchains_DMRG_exp,bera_spinon_2017,Gannon2019,wu2019,tran_spinon_2020,gao_spinon_2024}. The spinon description of the dynamical structure factor extends beyond the low-energy long-wavelength limit\cite{lake_finiteT_spinchains_DMRG_exp}. Spinon and holon excitations are also present in fermion systems\cite{moreno_nonlinear_2016}, and holon contributions are expected to dominate the momentum distribution\cite{tsyplyatyev_splitting_2022} near $3k_F$.

\subsection{Majorana and parafermionic quasiparticles in one-dimensional edge states}

Just a few years after the discovery of the QSH state by Kane and Mele \cite{KaneMele}, Fu and Kane \cite{FuKane2008, FuKane2009} had suggested that the QSH helical edge could potentially serve as a potential host platform for Majorana fermions when the edge is proximitized by a conventional $s$-wave superconductor \cite{Alicea2012}. Different from the early non-interacting modes, strongly-interacting QSH edge liquids \cite{hsu_helical_2021} promise to realize another type of non-Abelian anyon -- a `fractionalized Majorana' or $\mathbb{Z}_4$ parafermion, which could be detected in a 8$\pi$-periodic fractional Josephson effect \cite{XhangKane2014_Josephson, Orth2015}. Similar to the much sought-after Majorana fermion \cite{Alicea2012} $\mathbb{Z}_4$ parafermions have been proposed as intrinsically decoherence-free carriers of topologically protected quantum information \cite{Alicea_para}. But a key advantage of $\mathbb{Z}_4$ parafermions over Majoranas could be that -- similar to the Read-Rezayi \cite{ReadRezayi} fractional quantum Hall state -- these parafermions could be be further engineered into non-Abelian Fibonacci anyons, which would -- at least theoretically -- allow the realization of a universal gate set of a topological quantum computer \cite{Alicea_para}. 

\subsection{Superclimbing dislocations in solid Helium-4} 
Superflow through a one-dimensional edge dislocation core in solid $^4$He gives rise to physics beyond the Tomonaga-Luttinger liquid\cite{yarmolinsky_2017}. The Hamiltonian inside the core is 
\begin{equation}
H=\int dx \left[\frac{\chi} 2 (\nabla \rho)^2 + \rho_s (\nabla \theta)^2 \right], 
\end{equation}
where $h(x)$ is the displacement of the dislocation in the direction orthogonal its Burgers vector and its axis, $\theta$ is the superfluid phase, $\chi$ depends on the shear modulus of the crystal, and $\rho_s$ is the superfluid stiffness. The density is $\rho(x)=h(x)/a^2$, where $a$ is the lattice spacing of the $^4$He crystal, and $[\theta(x),h(x')]=i a^2 \delta(x-x')$.    In contrast to the Tomonaga-Luttinger liquid, translational invariance of the elastic Hamiltonian precludes the presence of  a $h^2 \sim \rho^2$ term, making the compressibility divergent. This allows both  excitations with quadratic instead of linear dispersion and long range superfluid order.\cite{yarmolinsky_2017,zhang_superclimbing_2024} The Peierls barrier $\propto \cos (2\pi h/a)$ can restore a nonzero compressibility and the Tomonaga-Luttinger liquid physics at long distances.\cite{yarmolinsky_2017}

\section{Outlook and  conclusions } \label{sec:conclusions}
The concept of the Tomonaga-Luttinger liquid (TLL) has proven to be an essential framework for understanding quasi one-dimensional (1D) quantum systems. Originally developed in connection with organic conductors, the TLL theory's broad applicability has been demonstrated across a wide variety of systems—from carbon nanotubes and quantum Hall / topological insulator edges to Josephson junctions and spin chains. The successful experimental realization of TLL physics in these diverse platforms not only confirmed the robustness of the theory but also broadened its relevance beyond fermionic systems to include also bosons and anyons. 
One of the most exciting developments in recent years has been the extension of TLL concepts to cold atom systems, where experimental control over parameters has allowed for unprecedented tests of the theory’s predictions. These platforms open new avenues for probing the limits of TLL physics and exploring quantum-criticality in 1D systems under various conditions, such as finite temperature\cite{hulet_2022}, finite size, engineered disorder\cite{aspect-dis-2008}, and tunable interactions.

We have discussed the power of conformal field theory to describe quantum-critical behaviors in one-dimensional systems. The interplay between collective excitations and absence of long-range order in 1D systems gives rise to a variety of fascinating phenomena, including absence of quasiparticle poles, spin charge separation and non-universal critical exponents.

Ultimately, the TLL model not only provides an accurate description of one-dimensional quantum systems but also continues to inspire new questions and explorations in condensed matter physics.

One major avenue lies in extending TLL physics beyond the traditional one-dimensional systems. The interplay between Luttinger liquid behavior and higher-dimensional systems presents an exciting opportunity to explore how TLL concepts may influence or coexist with other many-body effects. For instance, understanding how Luttinger liquid physics manifests itself in hinge states of higher-order topological insulators\cite{Schindler_2018,Wang2023}, layer domain walls in Van der Waals heterostructures\cite{li_2024}, carbon nanotubes deposited on graphene substrates\cite{flebus_electronic_2020} or atomic wires on seminconducting surfaces\cite{pfnur_atomic_2024} is an open question that may yield new quantum phases or exotic boundary states. Quite generally the question of the dimensional crossover arising when many one dimensional structures are coupled together, and how much of the peculiar one dimensional physics influences the behavior of the larger dimensional system is a considerable challenge and a route to tackle physics in higher dimensions. Challenges both on the theoretical and experimental side exist for spin, bosonic\cite{guo2023} and ultimately fermionic systems\cite{giamarchi_framework_2004,jerome_quasi_2024}. 

Moreover, the field stands to benefit greatly from continued advances in cold atom experiments. The extreme tunability of these platforms enables precise control over interactions, dimensionality, and particle statistics, providing a powerful testbed for exploring quantum-critical behavior, out-of-equilibrium dynamics, and disorder. The ability to simulate interacting Luttinger liquids in optical lattices or confined geometries could deliver fresh insights into the behavior of TLLs under conditions that were previously difficult or impossible to achieve in solid-state systems\cite{dipolar_2007}.

Additionally, the push toward studying non-equilibrium and dissipative dynamics in one-dimensional systems—where TLLs are expected to play a crucial role—opens up exciting new possibilities. Investigating how Luttinger liquids respond to time-dependent perturbations\cite{gutman_bosonization_2010,protopopov_many-particle_2011} or dissipation\cite{bacsi_dissipation-induced_2020,rosso_dynamical_2023,friedman_dissipative_2019,majumdar_bath-induced_2023} could reveal novel phenomena and quantum information processing applications. These questions will require the development of new theoretical tools to handle strongly interacting dissipative systems far from equilibrium, potentially leading to new paradigms in the study of quantum dynamics.

The application of TLL theory to more complex systems, such as edge modes in interacting topological phases,\cite{Kang2024,Xu2023,Cai2023,Park2023,Que2023} is another frontier. The growing intersection of TLL physics with quantum spin liquids\cite{poilblanc_2016} and topological matter\cite{thorngren_gapless_SPT_2021} suggests that there is still much to be learned about strongly correlated phases in reduced dimensions and bulk-boundary correspondence.

In conclusion, while TLL theory has already achieved remarkable success in explaining the behavior of 1D systems, its full potential is far from exhausted. Future research will reveal new connections between TLL physics and other areas of condensed matter, keeping it at the forefront of theoretical and experimental explorations in the quantum realm.

%

\section*{Acknowledgements}
R.C. was partly supported by the PNRR MUR project PE0000023 NQSTI (TOPQIN and SPUNTO). I.B., R.C., E.O. and T. G. would like to thank the Institut Henri Poincaré (UAR 839 CNRS Sorbonne Université) and the Lab Ex CARMIN(ANR-10 LABX-59-01)for their support. T. G. was supported in part by the Swiss National Science Foundation under grants 200020-188687 and 200020-219400. B. W. acknowledges the support of the National Research Foundation (NRF) Singapore, under the Competitive Research Program “Towards On-Chip Topological Quantum Devices” (NRF-CRP21-2018-0001), with further support from the Singapore Ministry of Education (MOE) Academic Research Fund Tier 3 grant (MOE-MOET32023-0003) “Quantum Geometric Advantage”. T. D. acknowledges support from the Centre of Excellence for Engineered Quantum Systems, an Australian Research Council Centre of Excellence, CE110001013. 
M. K. acknowledges the financial support of the Slovenian Research and Innovation Agency through the programme No. P1-0125 and the project No. J1-2456. RGH was funded in part by the National Science Foundation (US), grant no. 2309362
\section*{Author contributions}
The authors contributed equally to all aspects of the article. 

\section*{Competing interests}
The authors declare that they have no competing interests. 

\end{document}